\documentclass[fleqn,usenatbib]{mnras}

\usepackage{newtxtext,newtxmath}

\usepackage[T1]{fontenc}

\DeclareRobustCommand{\VAN}[3]{#2}
\let\VANthebibliography\thebibliography
\def\thebibliography{\DeclareRobustCommand{\VAN}[3]{##3}\VANthebibliography}

\usepackage{graphicx}	
\usepackage{amsmath}	

\usepackage{xcolor}

\title[Retrieval biases due to photospheric radius]{Impact of Variable Photospheric Radius on Exoplanet Atmospheric Retrievals}

\author[J Taylor]{
Jake Taylor$^{1}$\thanks{E-mail: jake.taylor@umontreal.ca}
\\
$^{1}$Institute for Research on Exoplanets, Department of Physics, University of Montréal, Montréal, H2V 0B3, Canada}

\date{Accepted XXX. Received YYY; in original form ZZZ}

\pubyear{2022}

\begin{document}
\label{firstpage}
\pagerange{\pageref{firstpage}--\pageref{lastpage}}
\maketitle

\begin{abstract}
Inverse techniques are used to extract information about an exoplanet's atmosphere. These techniques are prone to biased results if the appropriate forward model is not used. One assumption used in a forward model is to assume that the radius of the planet is constant with wavelength, however a more realistic assumption is that the photospheric radius varies with each wavelength. We explore the bias induced when attempting to extract the molecular abundance from an emission spectrum which was generated with a variable radius. We find that for low gravity planets, the retrieval model is not able to fit the data if a constant radius model is used. We find that biased results are obtained when studying a typical hot Jupiter in the MIRI LRS wavelength range. Finally, we show that high gravity planets do not suffer a bias. We recommend that future spectral retrievals that interpret exoplanet emission spectra should take into account a variable radius.
\end{abstract}




\section{Introduction} \label{sec:intro}
Studying the emission spectra of exoplanets using spectral retrievals has been the pioneering technique over the past decade to infer planetary properties \citep{Lee2012,Line2014b,Stevenson2014a,Line2016,Evans2017,Edwards2020,Changeat2021}. Current work has utilised both the Hubble Space Telescope (HST) and the Spitzer Space Telescope. With such limited spectral coverage, assumptions about the forward model used in the retrieval framework need to be made. These assumptions include, the parameterization of the temperature-pressure structure, the molecules in the atmosphere, and the geometry and the cloud composition.

With the successful launch of the James Webb Space Telescope (JWST) on the 25th of December 2021 the exoplanet community will have access to spectral observations of exoplanets from $0.6-12$ microns at unprecedented spectral resolution. The higher signal to noise observations will require more complex forward models such as geometries that extend past the current 1D assumption \citep{Feng2016,Taylor2020,Feng2020,Irwin2020,Changeat2020} and the parameterisation of clouds \citep{Taylor2021}. 

A common model assumption is that the radius of the planet is constant with wavelength. \citet{Fortney2019} compared the effect of computing the radius at each wavelength has on the emission spectra. They found that while the effect is not important for current HST observations, it can cause a $10-20$ \% difference in the wavelength dependent eclipse depth for JWST quality data. Therefore in this study we explore how the variable radius assumption can bias our retrieved results when studying JWST quality observations.

\section{Methods} \label{sec:methods}
In this section we outline the model set up and retrieval framework used.

The outline of the methodology of this study are as follows:
\begin{itemize}
    \item Generate a model of a hot Jupiter which has a radius that varies with wavelength
    \item Consider 3 different observing scenarios: NIRSpec PRISM, MIRI LRS and NIRSpec PRISM + MIRI LRS
    \item For each of the observing scenarios, considering 3 different error envelopes: 100ppm, 60ppm and 30ppm.
    \item Retrieve on these simulated datasets with a retrieval set up which assumes that the radius is fixed at each wavelength
\end{itemize}

\subsection{Model Setup}

\begin{figure*}
    \centering
    \includegraphics[width=\textwidth]{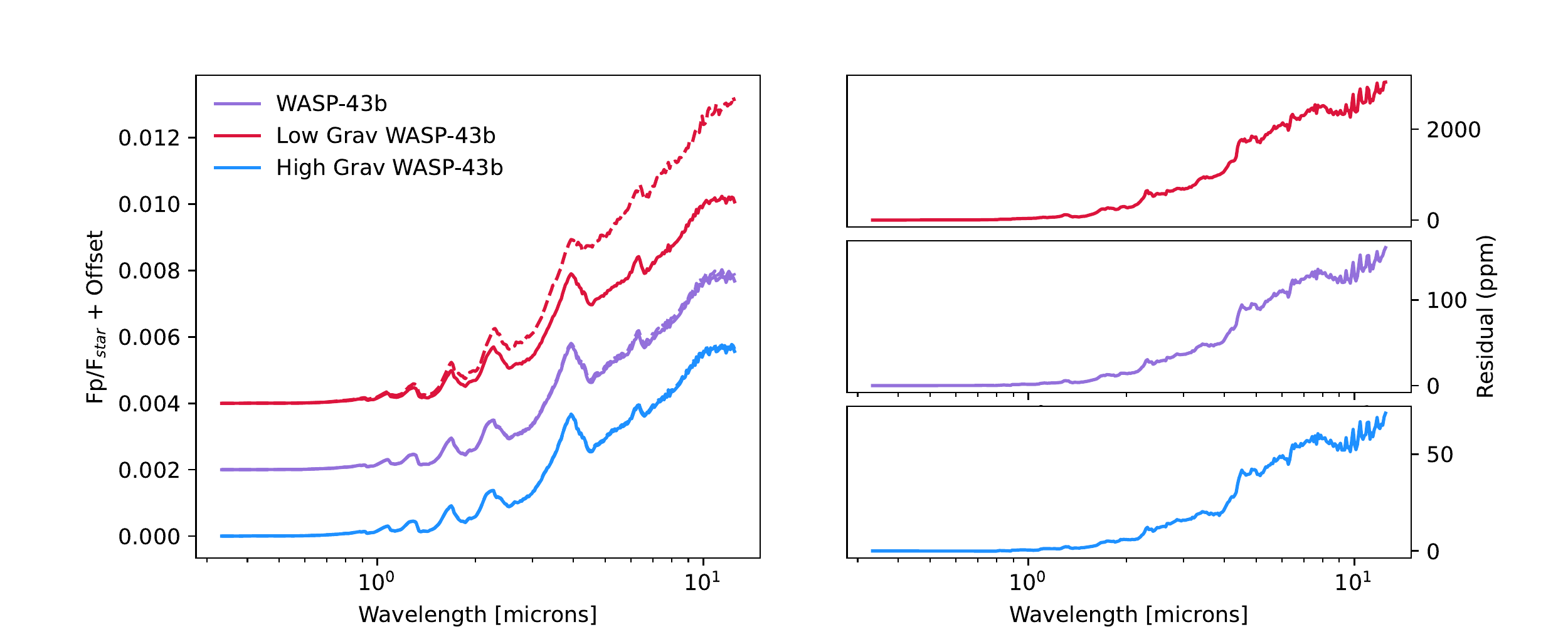}
    \caption{Left: Forward models used in this study with variable radius shown by the dashed line and constant radius with the solid line. In purple we present the model based on WASP-43b. In red we present the low gravity scenario. In blue we present the high gravity scenario. Right: The residuals when taking the difference between the variable and constant radius models. }
    \label{fig:forward_models}
\end{figure*}

It was found by \citet{Fortney2019} that the impact of the variable radius is dependent on the surface gravity of the planet. Hence, we explore 3 atmospheric scenarios: 
\begin{enumerate}
    \item A cloud free model based on the physical properties of WASP-43b.
    \item Same as (1) however with 10\% of the mass. We call this the low gravity scenario. This is analogous to a low gravity planet such as WASP-107b.
    \item Same as (1) however with 2$\times$ the mass. We call this the high gravity scenario. This is analogous to a high gravity planet such as WASP-18b.
\end{enumerate}
These scenarios let us explore a range of planetary scenarios observable with JWST. To generate these model spectra we use the CHIMERA code (see \citet{Line2013} for more detailed explanation of the capabilities of the framework, the open source code can be found here: \url{https://github.com/mrline/CHIMERA}), we show the main system parameters used in this study in Table \ref{tab:model_params}. We use a parameterised temperature-pressure profile developed by \citet{Parmentier2014} which balances incoming shortwave radiation with the outgoing longwave radiation. For simplicity we have used a blackbody spectrum calculated at T = 4400K for the stellar model. We present the models in Figure \ref{fig:forward_models}, it can see seen that when comparing the difference between a model with a variable radius and one with a constant radius, that there is a difference which increases with wavelength. We see that the difference is larger for lower gravity cases, which is consistent with \citet{Fortney2019}. We note that our study only considers hot Jupiters. \citet{Fortney2019} points out that the impact may be different for ultra hot Jupiters.

\begin{table}
   \centering
\begin{tabular}{l|l|l|l}
    \hline
    Parameter      & Value & Parameter      & Value   \\
    \hline
    log(H$_2$O)$^1$     & -3 & log(CO$_2$)$^4$     & -8 \\
    log(CO)$^2$     & -3 & log(CH$_4$)$^3$     & -8\\
    Teq [K]     & 1400& Rs [R$_{\text{sun}}$]$^5$      & 0.667 \\
    log($\kappa_{IR}$) & -1.5 & Tstar [K]     & 4400\\
    log($\gamma$) & -0.7 & SMA [AU]$^5$     & 0.015 \\
    Mp [M$_j$]$^5$     & 2.034 & Rp [R$_j$]$^5$     & 1.036
\end{tabular}
\caption{Atmospheric and planetary properties used to create the model atmospheres. $^1$\citet{polyansky2018exomol}, $^2$\citet{li2015rovibrational}, $^3$\citet{yurchenko2014exomol}, $^4$\citet{Yurchenko2020}, $^5$\citet{Gillon2012}.}
    \label{tab:model_params}
\end{table}

\subsection{Variable Radius Implementation}
\label{implementation}
We upgraded the CHIMERA code to be able to calculate the flux when considering a change in the photospheric radius (defined as the radius at which optical depth $\tau(\lambda)=2/3$) at each wavelength.

The planet-to-star flux is given by the equation:
\begin{equation}
\label{Eq1}
    F = \Big(\frac{R_p(\lambda)}{R_s(\lambda)}\Big)^2\frac{F_p(\lambda)d\lambda}{F_s(\lambda)d\lambda}
\end{equation}
where $R_p(\lambda)$ is the planetary radius, which is shown to have a wavelength dependence. The planetary radius is often assumed to be constant with wavelength.

To account for the wavelength dependence we rewrite Equation \ref{Eq1} to be:
\begin{equation}
    F = \Big(\frac{R_p + dR_p(\lambda)}{R_s(\lambda)}\Big)^2\frac{F_p(\lambda)d\lambda}{F_s(\lambda)d\lambda}   
\end{equation}
where $dR_p(\lambda)$ is the change in radius at each wavelength. We can compute this $dR_p(\lambda)$ by first extracting the altitude levels in the atmosphere ($Z(\lambda)$) and dividing this by the planetary radius ($R_0$), $Z(\lambda)/R_0$. By calculating the cumulative optical depth ($\tau(\lambda)$) as a function of $Z(\lambda)$, it is possible to determine the altitude at $\tau(\lambda) = 2/3$, hence this would provide the $dR_p(\lambda)$ at each wavelength. We note that in this study we are considering this effect in 1D and the $\tau(\lambda) = 2/3$ should vary for different emission angles.

\subsection{Sampling Method}
To perform the retrievals we used the dynesty package \citep{Speagle2020} which implements a nested sampling algorithm \citep{Skilling2004,Skilling2006}. This technique has been effective in extracting atmospheric information from spectra \citep{Rathcke2021,Lustig2021,Ahrer2022}.
\section{Results} \label{sec:results}
In this section we will outline the results from the three different planetary scenarios. We retrieve on the variable radius models shown in Figure \ref{fig:forward_models} using a fixed radius assumption in our retrieval model. We perform this for three different wavelength ranges obtainable with JWST. These are: NIRSpec PRISM, MIRI LRS and then combining the two instrument modes. For each of these wavelength ranges a 100ppm, 60ppm and 30ppm error envelope is assumed to assess the limits in which a bias may arise.

\subsection{WASP-43b}
\begin{figure}
    \centering
    \includegraphics[width=0.45\textwidth]{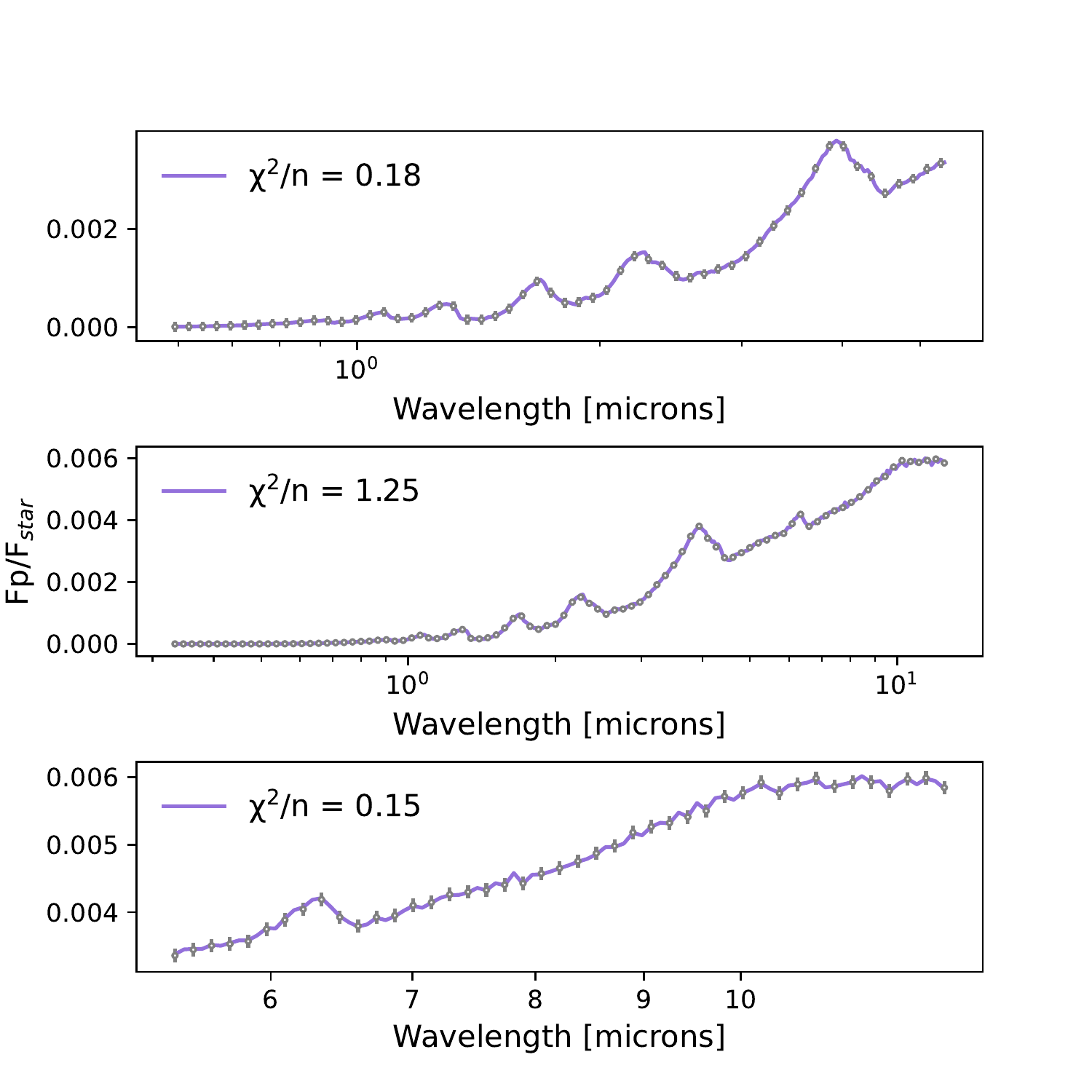}
    \caption{The best fitting model and simulated data for WASP-43b. The top, middle, and bottom panel show the NIRSpec observing mode, the NIRSpec + MIRI LRS observing mode, the MIRI LRS observing mode, respectively. The reduced $\chi^2$ of the fit is shown inset.}
    \label{fig:WASP-43b_Spectra}
\end{figure}

Figure \ref{fig:forward_models} shows that the difference between the variable and constant radius models do not deviate until mid-IR, with around 100ppm difference at 4.5 microns. Figure \ref{fig:WASP-43b_Spectra} shows the best fitting spectra for all the observing modes for the 100ppm case. Table \ref{tab:WASP-43b_table} shows the retrieved abundances for each modelling set up, accompanied by the reduced $\chi^2$.

Observations with NIRSpec PRISM show that the variable radius does not bias the retrieved chemical abundances.
For both the 100ppm and 60ppm cases, the abundance of CO$_2$ is unconstrained, however at the 30ppm level it is possible to constrain the abundance of CO$_2$. For observations with MIRI LRS, the abundance of H$_2$O is retrieved to the correct value within 2$\sigma$ for the 100ppm and 60ppm case. A biased result is retrieved for the 30ppm case. When combining instrument modes H$_2$O and CO are retrieved to be greater than 2$\sigma$ away from the input value. It is interesting that the combining NIRSpec PRISM with MIRI LRS results in a biased abundance of CO, as there is no CO information in this wavelength region. When combining instrument modes we begin to detect CO$_2$ at the 60ppm level and then constrain the abundance to the correct value at the 30ppm level. 

\subsection{Low Gravity Scenario}
\begin{figure}
    \centering
    \includegraphics[width=0.45\textwidth]{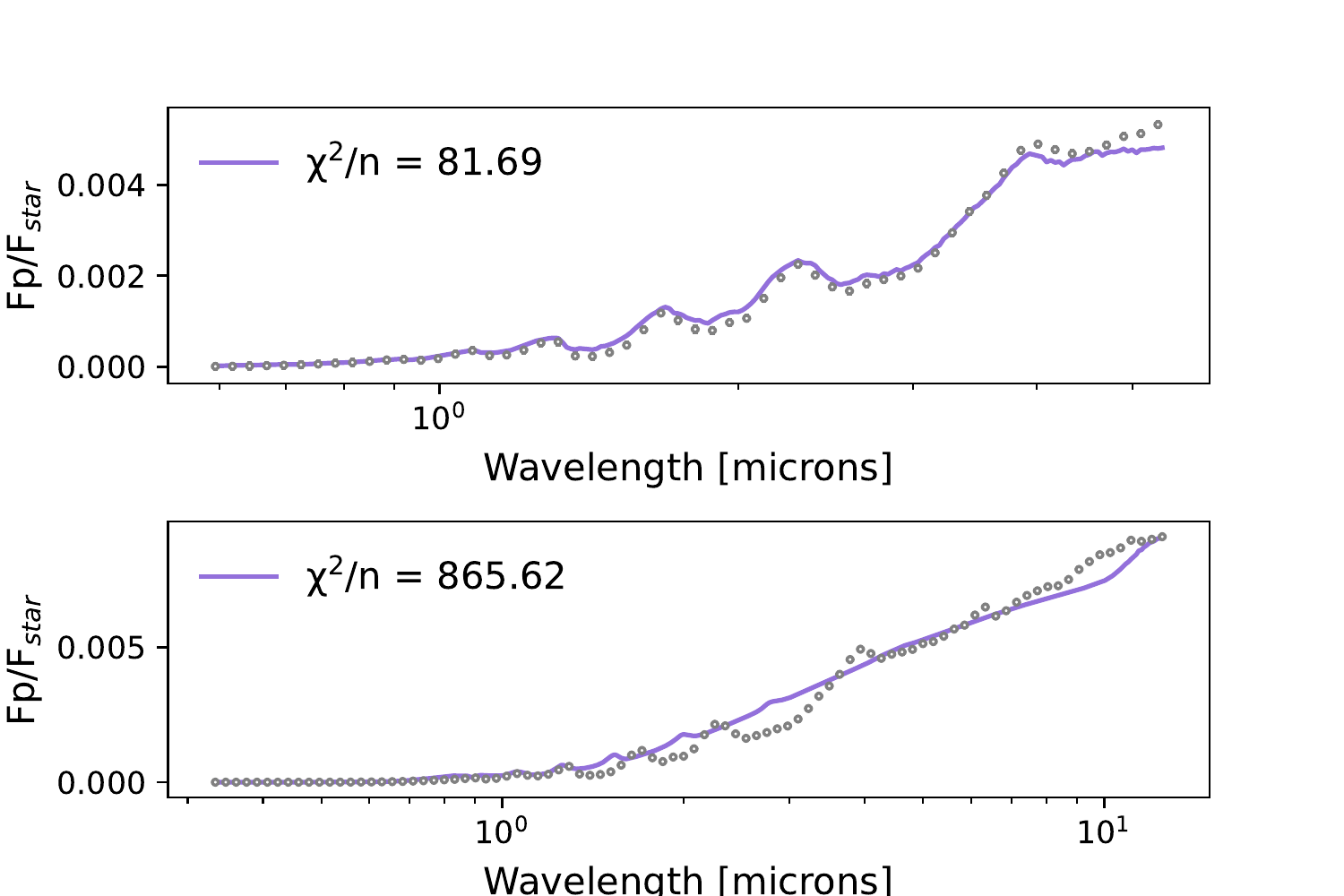}
    \caption{The best fitting model and simulated data for the low gravity case. The top and bottom panel show the NIRSpec observing mode and the NIRSpec + MIRI LRS observing mode, respectively. The reduced $\chi^2$ of the fit is shown inset.}
    \label{fig:Low_Grav_Spectra}
\end{figure}

Figure \ref{fig:forward_models} shows that the difference between the variable and constant radius models is large throughout the whole infrared. Although the initial 100ppm simulation produced the incorrect chemistry, this is not the interesting takeaway point.
The constant radius model is not able to fit the variable radius simulated data, we show this in Figure \ref{fig:Low_Grav_Spectra}. Therefore low gravity planets will require a variable radius model to be able to fit the observations. After 17 days the MIRI LRS still had not converged, compared to 0.477 days and 0.282 days from the WASP-43b and High Gravity cases. Each simulation was run on 12 central processing units (CPU) cores.
The forward model, which assumes a constant radius, is too simple. The sampling algorithm cannot converge to produce an adequate reduced $\chi^2$. As a result of the biases being observed at the 100ppm level, we do not explore the 60ppm or 30ppm level.

\subsection{High Gravity Scenario}
\begin{figure}
    \centering
    \includegraphics[width=0.45\textwidth]{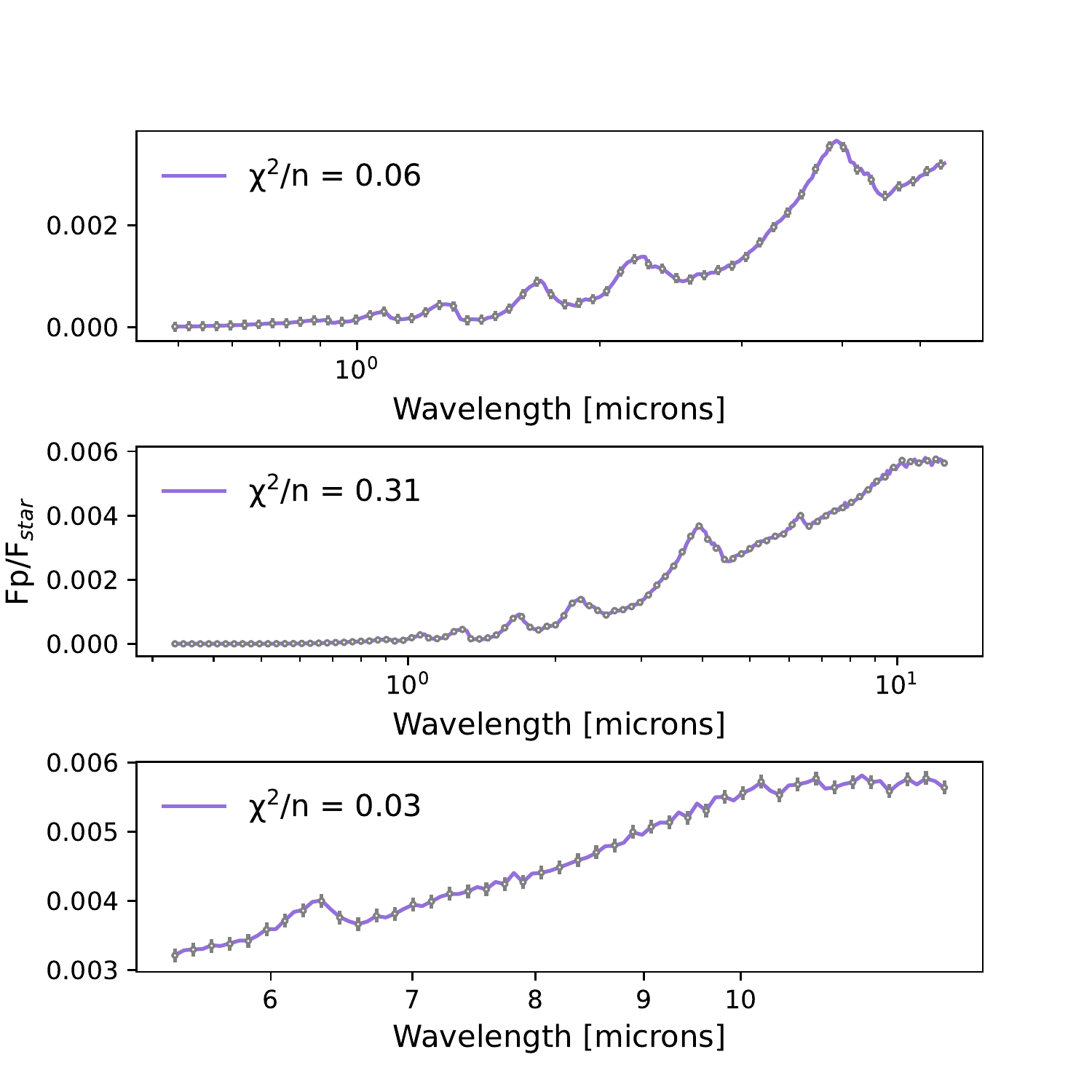}
    \caption{The best fitting model and simulated data for the high gravity case. The top, middle, and bottom panel show the NIRSpec observing mode, the NIRSpec + MIRI LRS observing mode, the MIRI LRS observing mode, respectively. The reduced $\chi^2$ of the fit is shown inset.}
    \label{fig:High_Grav_Spectra}
\end{figure}

Figure \ref{fig:forward_models} shows that the difference in the models is only around 50ppm at 4.5 microns. Figure \ref{fig:High_Grav_Spectra} shows the best fitting models from the 100ppm case and present the retrieved abundances in Table \ref{tab:High_Grav_table}. For each of the NIRSpec PRISM and NIRSpec PRISM + MIRI LRS observing modes, the abundance of H$_2$O and CO were constrained within 2$\sigma$ of the input value. For the MIRI LRS observing mode the H$_2$O abundance was constrained within 2$\sigma$ and the rest of the chemistry is unconstrained, which was to be expected. We can summarise that a constant radius model is sufficient to interpret the atmosphere of a high gravity exoplanet.

\section{Conclusion} \label{sec:conclusion}
In this study we explored the impact of a variable photospheric radius on a retrieval analysis of emission spectra.
We explored three model scenarios which span a range of gravities. Our conclusions can be summarised as follows:
\begin{enumerate}
    \item For a typical hot Jupiter, the variable radius assumption is not important when interpreting observations in the NIRSpec PRISM wavelength range. However, the impact of a variable radius becomes important when observing over the MIRI LRS wavelength region. If the variable radius is not assumed in the retrieval model then biased chemistry will be found.
    \item For a low gravity hot Jupiter it is not possible to fit the emission spectrum with a constant radius model. Hence for all observing scenarios of a low gravity hot Jupiter, a variable radius model needs to be used for accurate interpretation.
    \item For a high gravity hot Jupiter no biases were found when retrieving on the variable radius model with a constant radius model. 
    \item It is possible to constrain the abundance of CO$_2$ unambiguously in the WASP-43b case with an error envelope of 30ppm. A similar constraint is placed on the abundance of CO$_2$ in the high gravity case. The upper limit is well constrained, however a large tail is seen in the lower limit.
\end{enumerate}

From this study we recommend that the community include a variable radius model in their retrieval frameworks as it will be important in the JWST era. This can be implemented in a similar way as discussed in Section \ref{implementation}. Given that the optical depths are already computed in the radiative transfer calculation, determining and then adding $dR_p(\lambda)$ does not increase computation time.

\section{Acknowledgements}
Jake Taylor thanks the Canadian Space Agency for financially supporting this work. He thanks Dr Peter McGill for proof reading the manuscript. Jake thanks Dr Michael Line for discussions about the photospheric variable radius and the use of CHIMERA. Jake thanks Dr Joanna Barstow for the review of this letter which greatly improved its clarity. 

\section*{Data Availability}

Any data generated in this study is available on request.



\bibliographystyle{mnras}
\bibliography{bib} 




\appendix

\begin{table*}
    \centering
    \begin{tabular}{cccccc}
    NIRSpec PRISM& & & & &\\
    \hline
    Model & logH$_2$O (-3) & logCO (-3) & logCO$_2$ (-8) & logCH$_4$ (-8) &$\chi^2$/n\\
    \hline
    100ppm     & \color{violet}{-2.99$^{+0.40}_{-0.24}$} &\color{violet}{-3.05$^{+0.63}_{-0.43}$} &-9.67$^{+2.10}_{-2.20}$ &-9.35$^{+2.08}_{-1.99}$ &0.18\\
    60ppm     & \color{violet}{-3.02$^{+0.19}_{-0.14}$} &\color{violet}{-3.10$^{+0.32}_{-0.26}$} & -9.68$^{+1.91}_{-2.21}$&-9.46$^{+2.56}_{-2.42}$ &0.51 \\
    30ppm     & \color{violet}{-3.03$^{+0.09}_{-0.07}$} &\color{violet}{-3.12$^{+0.14}_{-0.13}$} &\color{blue}{-8.74$^{+0.85}_{-3.10}$} &-9.59$^{+2.35}_{-2.29}$ &1.89\\
    \hline
    MIRI LRS& & & & &\\
    \hline
    Model & logH$_2$O (-3) & logCO (-3) & logCO$_2$ (-8) & logCH$_4$ (-8) &$\chi^2$/n\\
    \hline
    100ppm     & \color{violet}{-2.24$^{+1.01}_{-1.56}$} & -6.65$^{+4.80}_{-5.05}$& -8.25$^{+3.49}_{-3.53}$& -8.68$^{+3.16}_{-3.10}$& 0.15 \\
    60ppm     & \color{violet}{-2.24$^{+0.80}_{-0.99}$} & -6.75$^{+4.90}_{-4.88}$&-8.60$^{+3.67}_{-3.23}$ & -8.81$^{+2.97}_{-3.01}$& 0.39 \\
    30ppm     & \color{red}{-2.21$^{+0.37}_{-0.38}$} & -6.75$^{+4.76}_{-4.92}$& -8.61$^{+3.16}_{-3.08}$& -9.09$^{+2.87}_{-2.69}$& 1.58\\
    \hline
    NIRSpec PRISM + MIRI LRS& & & & &\\
    \hline
    Model & logH$_2$O (-3)& logCO (-3)& logCO$_2$ (-8)& logCH$_4$ (-8)&$\chi^2$/n\\
    \hline
    100ppm     & \color{violet}{-2.68$^{+0.38}_{-0.33}$} & \color{red}{-2.52$^{+0.44}_{-0.42}$}& -9.20$^{+2.05}_{-2.65}$& -9.03$^{+2.67}_{-2.81}$& 1.25 \\
    60ppm     & \color{red}{-2.68$^{+0.22}_{-0.21}$} & \color{red}{-2.53$^{+0.26}_{-0.27}$}&\color{blue}{-8.29$^{+1.03}_{-3.51}$} & -9.19$^{+2.61}_{-2.67}$& 3.43 \\
    30ppm     & \color{red}{-2.68$^{+0.11}_{-0.11}$} & \color{red}{-2.54$^{+0.14}_{-0.14}$}& \color{violet}{-7.60$^{+0.22}_{-0.41}$}& -9.13$^{+2.41}_{-2.71}$& 12.24\\ 
    \end{tabular}
    \caption{The retrieved chemical abundances for the WASP-43b model. We show each observing mode and error envelope. We also present the reduced $\chi^2$ where n = 7. The error ranges are the 2$\sigma$ ranges as outputted from dynesty. We present the true model values in brackets next to the parameters name. We have colour coordinated the results so they are easier for the reader to interpret. We present results within 2$\sigma$ in purple, outside of 2$\sigma$ in red, results with an upper limit inside 2$\sigma$ in blue and unconstrained results in black.} 
    \label{tab:WASP-43b_table}
\end{table*}

\begin{table*}
    \centering
    \begin{tabular}{cccccc}
    NIRSpec PRISM& & & & &\\
    \hline
    Model & logH$_2$O (-3)& logCO (-3)& logCO$_2$ (-8)& logCH$_4$ (-8)&$\chi^2$/n\\
    \hline
    100ppm     & \color{red}{-2.00$^{+0.18}_{-0.25}$} &-8.16$^{+3.81}_{-3.66}$ &-9.93$^{+2.28}_{-1.96}$ &-8.77$^{+3.18}_{-3.05}$ &81.96\\
    \hline
    NIRSpec PRISM + MIRI LRS& & & & &\\
    \hline
    Model & logH$_2$O (-3)& logCO (-3)& logCO$_2$ (-8)& logCH$_4$ (-8)&$\chi^2$/n\\
    \hline
    100ppm     & -9.26$^{+2.65}_{-2.57}$ & \color{violet}{-1.76$^{+0.55}_{-2.14}$}& \color{blue}{-7.03$^{+1.54}_{-4.61}$}& \color{red}{-1.83$^{+0.28}_{-0.27}$}& 865.62 \\
    \end{tabular}
    \caption{The retrieved chemical abundances for the Low Gravity model. We show each observing mode and error envelope. We also present the reduced $\chi^2$ where n = 7. The error ranges are the 2$\sigma$ ranges as outputted from dynesty.We present the true model values in brackets next to the parameters name. We have colour coordinated the results so they are easier for the reader to interpret. We present results within 2$\sigma$ in purple, outside of 2$\sigma$ in red, results with an upper limit inside 2$\sigma$ in blue and unconstrained results in black.} 
    \label{tab:Low_Grav_table}
\end{table*}

\begin{table*}
    \centering
    \begin{tabular}{cccccc}
    NIRSpec PRISM& & & & &\\
    \hline
    Model & logH$_2$O (-3)& logCO (-3)& logCO$_2$ (-8)& logCH$_4$ (-8)&$\chi^2$/n\\
    \hline
    100ppm     & \color{violet}{-2.99$^{+0.30}_{-0.20}$} &\color{violet}{-3.01$^{+0.50}_{-0.42}$} &-9.61$^{+2.05}_{-2.28}$ &-9.29$^{+2.75}_{-2.55}$ &0.06\\
    60ppm     & \color{violet}{-3.01$^{+0.15}_{-0.13}$} &\color{violet}{-3.04$^{+0.28}_{-0.25}$} & -9.62$^{+1.94}_{-2.25}$&-9.37$^{+2.53}_{-2.50}$ &0.15 \\
    30ppm     & \color{violet}{-3.01$^{+0.07}_{-0.07}$} &\color{violet}{-3.05$^{+0.13}_{-0.13}$} &-9.01$^{+1.18}_{-2.82}$ &-9.54$^{+2.41}_{-2.35}$ &0.50\\
    \hline
    MIRI LRS& & & & &\\
    \hline
    Model & logH$_2$O (-3)& logCO (-3)& logCO$_2$ (-8)& logCH$_4$ (-8)&$\chi^2$/n\\
    \hline
    100ppm     & \color{violet}{-2.47$^{+1.26}_{-1.40}$} & -6.79$^{+5.09}_{-4.90}$& -8.40$^{+3.70}_{-3.39}$& -8.74$^{+3.28}_{-3.08}$& 0.03 \\
    60ppm     & \color{violet}{-2.50$^{+0.95}_{-1.21}$} & -6.84$^{+4.89}_{-4.73}$&-8.19$^{+3.31}_{-3.50}$ & -8.84$^{+3.06}_{-2.92}$& 0.10 \\
    30ppm     & \color{violet}{-2.49$^{+0.50}_{-0.69}$} & -6.84$^{+4.81}_{-4.80}$& -8.64$^{+3.28}_{-3.07}$& -9.10$^{+3.02}_{-2.72}$& 0.39\\
    \hline
    NIRSpec PRISM + MIRI LRS& & & & &\\
    \hline
    Model & logH$_2$O (-3)& logCO (-3)& logCO$_2$ (-8)& logCH$_4$ (-8)&$\chi^2$/n\\
    \hline
    100ppm     & \color{violet}{-2.93$^{+0.30}_{-0.22}$} & \color{violet}{-2.87$^{+0.39}_{-0.35}$}& -9.59$^{+2.20}_{-2.29}$& -9.06$^{+2.61}_{-2.81}$& 0.31 \\
    60ppm     & \color{violet}{-2.95$^{+0.15}_{-0.14}$} & \color{violet}{-2.89$^{+0.22}_{-0.21}$}&-9.39$^{+1.81}_{-2.48}$ & -9.14$^{+2.53}_{-2.71}$& 0.91 \\
    30ppm     & \color{violet}{-2.96$^{+0.07}_{-0.07}$} & \color{violet}{-2.90$^{+0.10}_{-0.10}$}& \color{blue}{-8.11$^{+0.44}_{-3.57}$}& -9.07$^{+2.30}_{-2.79}$& 3.29\\ 
    \end{tabular}
    \caption{The retrieved chemical abundances for the High Gravity model. We show each observing mode and error envelope. We also present the reduced $\chi^2$ where n = 7. The error ranges are the 2$\sigma$ ranges as outputted from dynesty. We present the true model values in brackets next to the parameters name. We have colour coordinated the results so they are easier for the reader to interpret. We present results within 2$\sigma$ in purple, outside of 2$\sigma$ in red, results with an upper limit inside 2$\sigma$ in blue and unconstrained results in black.}
    \label{tab:High_Grav_table}
\end{table*}


\bsp	
\label{lastpage}
\end{document}